\documentstyle[12pt]{article}

\begin{document}

\begin{titlepage}
\setcounter{page}{1}

\author{Waldemar Puszkarz\thanks{
Electronic address: puszkarz@cosm.sc.edu} \\
{\small {\it Department of Physics and Astronomy,} }\\
{\small {\it University of South Carolina,} }\\
{\small {\it Columbia, SC 29208}}}
\title{{\bf Extension of the Staruszkiewicz Modification of the 
Schr\"{o}dinger Equation}}
\date{{\small (May 15, 1999)}}
\maketitle

\begin{abstract}
We present an extension of Staruszkiewicz's modification of the
Schr\"{o}dinger equation which preserves its main and unique feature: in the 
natural system of units the modification terms do not contain any dimensional 
constants. The extension, similarly as the original, is formulated in a 
three-dimensional space and derives from a Galilean invariant Lagrangian. 
It is pointed out that this model of nonlinearity violates the separability 
of compound systems in the fundamentalist approach to this issue. In its 
general form, this modification does not admit stationary states for all 
potentials for which such states exist in linear quantum mechanics. 
This is, however, possible for a suitable choice of its free parameters. 
It is only in the original Staruszkiewicz modification that the energy of 
these states remains unchanged, which marks the uniqueness of this 
variant of the modification.

\vskip 0.5cm \noindent

\end{abstract}
\end{titlepage}


Several years ago Staruszkiewicz \cite{Sta1} proposed unique for three
dimensions a modification of the fundamental equation of quantum mechanics
by suggesting that the term $\gamma (\Delta S)^{2}$ be added to its action
density. Here and in what follows $S$ stands for the phase of the wave
function\footnote{%
It should be noted that our convention here is different from the standard
convention that employes the Planck constant in the exponent so that $S$ has
the dimensions of action. Here $S$ is dimensionless for it represents the
angle.} $\Psi =R\exp \left( iS\right) $. The unique character of this
modification stems from the fact that the constant $\gamma $ is
dimensionless in the system of natural units $\hbar =c=1$ in three
dimensions. This is unlike in any other modification put forward so far, the
most notable examples being the modification by Bia\l ynicki-Birula and
Mycielski \cite{Bial}, a very general modification of Weinberg \cite{Wein1},
and recently studied very extensively the modification of Doebner and Goldin 
\cite{Doeb}. Each of these modifications introduces some dimensional
parameters, the last of them in the total number of six. The Staruszkiewicz
modification, similarly as the modifications of Weinberg and Bia\l
ynicki-Birula and Mycielski, is Galilean invariant for an arbitrary value of
its coupling constant. This is a simple consequence of the fact that it
derives from a Galilean invariant Lagrangian. For comparison, the
Doebner-Goldin modification preserves the Galilean invariance only for a
certain selection of its six free parameters.

Out of the modifications mentioned, only the Bia\l ynicki-Birula and
Mycielski and Weinberg ones have been confronted with some sort of
experimental data and thereby upper bounds on their parameters have been
imposed. It seems that the former, to quote \cite{Bial2}, ``has been
practically ruled out by extremely accurate measurements of neutron
diffraction on an edge.'' The measurements in question \cite{Gahl} (see also 
\cite{Klein, Zeil} for a more comprehensive discussion of these and other
relevant techniques and experiments) have established an upper limit on the
only nonlinear parameter of the modification to be $3.3\times 10^{-15}$ eV.
The Weinberg general scheme leads to a number of particular models of
nonlinear Schr\"{o}dinger equation. Several experiments \cite{Boll, Chup,
Wal, Maj}, based on different ideas and using different techniques, have
been performed to test the most basic of these models in some reasonably
simple physical situations as, for instance, in the case of transitions
between two atomic levels \cite{Wein2, Wein1}. The data gathered in these
experiments imply an upper limit for the nonlinearity parameter of the
models to be of the order of $|1\times 10^{-20}|$ eV \cite{Maj, Boll} which
is the most stringent upper bound on nonlinear corrections to quantum
mechanics yet to date.

It is the purpose of the present report to show that the Staruszkiewicz
modification can be even further extended preserving its main features just
elaborated on. We will formulate the extension for a three-dimensional
space. In fact, as we will see, it is in this dimension that the proposed
extension results in much greater complexity than it is the case for lower
dimensions. To put our work in a proper perspective, let us first briefly
recall the main points of the Staruszkiewicz modification after his original
paper \cite{Sta1}. In what follows we will retain $\hbar $.

When the Lagrangian density for the Schr\"{o}dinger equation, 
\begin{eqnarray}
L_{SE} &=&\frac{i\hbar }{2}\left( \Psi ^{*}\frac{\partial \Psi }{\partial t}%
-\Psi \frac{\partial \Psi ^{*}}{\partial t}\right) -\frac{\hbar ^{2}}{2m}%
\left( \vec{\nabla}\Psi \right) ^{^{*}}\cdot \vec{\nabla}\Psi -V\Psi \Psi
^{*}=  \nonumber \\
&&-\left( \hbar R^{2}\frac{\partial S}{\partial t}+\frac{\hbar ^{2}}{2m}%
\left[ \left( \vec{\nabla}R\right) ^{2}+R^{2}\left( \vec{\nabla}S\right)
^{2}\right] +R^{2}V\right) ,  \label{1}
\end{eqnarray}
is supplemented by the above mentioned term with a suitably chosen numerical
factor in front of $\gamma $, the modified action for the Schr\"{o}dinger
equation in the hydrodynamic formulation, 
\begin{equation}
S_{mod}({\vec{r}},t)=-\int dt\,d^{3}x\left\{ \hbar R^{2}\frac{\partial S}{%
\partial t}+\frac{\hbar ^{2}}{2m}\left[ \left( \vec{\nabla}R\right)
^{2}+R^{2}\left( \vec{\nabla}S\right) ^{2}\right] +\frac{\gamma }{2}\left(
\Delta S\right) ^{2}+R^{2}V\right\} ,  \label{2}
\end{equation}
leads to two equations 
\begin{equation}
\hbar \frac{\partial R^{2}}{\partial t}+\frac{\hbar ^{2}}{m}\vec{\nabla}%
\cdot \left( R^{2}\vec{\nabla}S\right) -\gamma \Delta \Delta S=0,  \label{3}
\end{equation}
\begin{equation}
\frac{\hbar ^{2}}{m}\Delta R-2\hbar R\frac{\partial S}{\partial t}-2RV-\frac{%
\hbar ^{2}}{m}R\left( \vec{\nabla}S\right) ^{2}=0.  \label{4}
\end{equation}
Of these, only the former is different for the linear Schr\"{o}dinger
equation in the same formulation. It is the continuity equation for the
current 
\begin{equation}
\vec{j}=\frac{\hbar ^{2}}{m}R^{2}\vec{\nabla}S-\gamma \vec{\nabla}\Delta S
\label{5}
\end{equation}
and the probability density $\rho =R^{2}$. Physically acceptable solutions
to (3) and (4) are those that are both normalizable, i.e., $\int
d^{3}x\,R^{2}=1$ and satisfy the condition of finite energy 
\begin{equation}
E=\int \,d^{3}x\,\left\{ \frac{\hbar ^{2}}{2m}\left[ \left( \vec{\nabla}%
R\right) ^{2}+R^{2}\left( \vec{\nabla}S\right) ^{2}\right] +\frac{\gamma }{2}%
\left( \Delta S\right) ^{2}+R^{2}V\right\} <\infty .  \label{6}
\end{equation}
It turns out that this condition discriminates possible solutions rather
severly. For instance, as found in \cite{Sta1}, ordinary Gaussian wave
packets, even if solutions to (3-4), do not meet it. This energy functional
is derived within the field-theoretical Lagrangian approach and can differ
from the quantum-mechanical one as discussed in \cite{Pusz1, Pusz2} and also
below.

One observes that $\Delta S=$ $\frac{i}{2}\Delta (ln\frac{\Psi ^{*}}{\Psi })$
and so is homogeneous of degree zero in $\Psi $ and $\Psi ^{*}$. This term
is Galilean invariant under the Galilean covariant transformation of the
phase, 
\begin{equation}
S^{\prime }(x^{\prime },t^{\prime })=S(x,t)-m\vec{v}\cdot \vec{x}+\frac{1}{2}%
m\vec{v}^{2}t.  \label{7}
\end{equation}
However, this is not the only term built from $\Psi $ and $\Psi ^{*}$ that
has these properties. Terms of the form $\nabla lna\Psi \Psi ^{*}$ or $%
\Delta lna\Psi \Psi ^{*}$, where $a$ is an auxiliary dimensional constant%
\footnote{%
This constant is employed only to make the argument of logarithm
dimensionless and does not appear in the equations of motions nor does it in
any other physically relevant expressions.}, fall into the same category and
thus can be used to construct the Lagrangian density for an extended version
of the Staruszkiewicz modification. It is straightforward to convince
oneself that in three dimensions this density is 
\begin{eqnarray}
L_{ext} &=&L_{SE}+b_{1}(\Delta S)^{2}+b_{2}\Delta S\Delta
lnaR^{2}+b_{3}\Delta S(\nabla lnaR^{2})^{2}  \nonumber \\
&&+b_{4}(\Delta lnaR^{2})^{2}+b_{5}\Delta lnaR^{2}(\nabla
lnaR^{2})^{2}+b_{6}(\nabla lnaR^{2})^{4}.  \label{8}
\end{eqnarray}
Some of the terms in this Lagrangian contain similar expressions. To avoid
repeating them we rewrite it as 
\begin{eqnarray}
L_{ext} &=&-\left( \hbar R^{2}\frac{\partial S}{\partial t}+\frac{\hbar ^{2}%
}{2m}\left[ \left( \vec{\nabla}R\right) ^{2}+R^{2}\left( \vec{\nabla}%
S\right) ^{2}\right] +R^{2}V\right) -c_{1}(\Delta S)^{2}-c_{2}\Delta S\frac{%
\Delta R}{R}  \nonumber \\
&&-c_{3}\Delta S\left( \frac{\vec{\nabla}R}{R}\right) ^{2}-c_{4}\left( \frac{%
\Delta R}{R}\right) ^{2}-c_{5}\frac{\Delta R}{R}\left( \frac{\vec{\nabla}R}{R%
}\right) ^{2}-c_{6}\left( \frac{\vec{\nabla}R}{R}\right) ^{4},  \label{9}
\end{eqnarray}
where $c_{1}=-b_{1},c_{2}=-2b_{2},c_{3}=2b_{2}-4b_{3},c_{4}=-4b_{4},
c_{5}=8b_{4}-8b_{5},c_{6}=8b_{5}-4b_{4}. $ The equations of motion, 
\begin{equation}
\frac{\partial L}{\partial \varphi _{i}}-\partial _{\nu }\left( \frac{%
\partial L}{\partial _{_{\nu }}\varphi _{i}}\right) +\partial _{\nu
}\partial _{\mu }\left( \frac{\partial L}{\partial _{_{\nu }}\partial _{\mu
}\varphi _{i}}\right) =0,  \label{10}
\end{equation}
(with $\varphi =(R,S),i=1,2$) derived from the discussed Lagrangian read 
\begin{equation}
\hbar \frac{\partial R^{2}}{\partial t}+\frac{\hbar ^{2}}{m}\vec{\nabla}%
\cdot \left( R^{2}\vec{\nabla}S\right) -H_{I}R^{2}=0  \label{11}
\end{equation}
and 
\begin{equation}
\frac{\hbar ^{2}}{m}\Delta R-2\hbar R\frac{\partial S}{\partial t}-\frac{%
\hbar ^{2}}{m}R\left( \vec{\nabla}S\right) ^{2}-2RV-H_{R}R=0,  \label{12}
\end{equation}
where 
\begin{equation}
H_{I}=\left[ c_{1}\Delta \Delta S+c_{2}\Delta \left( \frac{\Delta R}{R}%
\right) +c_{3}\Delta \left( \frac{\vec{\nabla}R}{R}\right) ^{2}\right] \frac{%
1}{R^{2}}=\frac{h_{I}}{R^{2}}  \label{13}
\end{equation}
and 
\begin{eqnarray}
H_{R} &=&\left\{ 2c_{2}\Delta \left( \frac{\Delta S}{R}\right) +2c_{4}\Delta
\left( \frac{\Delta R}{R^{2}}\right) +c_{5}\Delta \left[ \frac{1}{R}\left( 
\frac{\vec{\nabla}R}{R}\right) ^{2}\right] -\right.  \nonumber \\
&&\vec{\nabla}\cdot \left[ 2c_{3}\left( \Delta S\frac{\vec{\nabla}R}{R^{2}}%
\right) +2c_{5}\left( \Delta R\frac{\vec{\nabla}R}{R^{3}}\right) +4c_{6}%
\frac{1}{R}\left( \frac{\vec{\nabla}R}{R}\right) ^{3}\right] -\frac{1}{R}%
\left[ c_{2}\Delta S\frac{\Delta R}{R}+\right.  \nonumber \\
&&\left. \left. 2c_{3}\Delta S\left( \frac{\vec{\nabla}R}{R}\right)
^{2}+2c_{4}\left( \frac{\Delta R}{R}\right) ^{2}+3c_{5}\frac{\Delta R}{R}%
\left( \frac{\vec{\nabla}R}{R}\right) ^{2}+4c_{6}\left( \frac{\vec{\nabla}R}{%
R}\right) ^{4}\right] \right\} \frac{1}{R}=\frac{h_{R}}{R^{2}}.  \label{14}
\end{eqnarray}
The coupling constants of the extended part of the Lagrangian are
dimensionless in the natural system of units in three dimensions.

One can replace these equations with a single equation for the wave function 
$\Psi $ 
\begin{equation}
i\hbar \frac{\partial \Psi }{\partial t}=\left( -\frac{\hbar ^{2}}{2m}\Delta
+V\right) \Psi +H_{NL}^{^{\prime }}\left[ \Psi ,\Psi ^{*}\right] \Psi ,
\label{15}
\end{equation}
where $H_{NL}^{^{\prime }}\left[ \Psi ,\Psi ^{*}\right] $ is a certain
functional of $\Psi $ and $\Psi ^{*}$. When expressed in terms of $R$ and $S$%
, it is simply 
\begin{equation}
H_{NL}^{^{\prime }}\left[ \Psi ,\Psi ^{*}\right] =\frac{1}{2}H_{R}+\frac{i}{2%
}H_{I}.  \label{16}
\end{equation}
The total Hamiltonian for the modification is 
\begin{equation}
H^{\prime }=H_{SE}+H_{NL}^{^{\prime }}=H_{R}^{^{\prime }}+H_{I}^{^{\prime }},
\label{17}
\end{equation}
where $H_{SE}$ stands for the Hamiltonian of the linear Schr\"{o}dinger
equation. In standard quantum mechanics, the mean value of this operator
represents the energy of a quantum-mechanical system and therefore one
expects that 
\begin{equation}
E_{QM}=\int \,d^{3}x\Psi ^{*}H^{\prime }\Psi =\int \,d^{3}x\left\{ \frac{%
\hbar ^{2}}{2m}\left[ \left( \vec{\nabla}R\right) ^{2}+R^{2}\left( \vec{%
\nabla}S\right) ^{2}\right] +VR^{2}+\frac{1}{2}h_{R}+\frac{i}{2}%
h_{I}\right\} .  \label{18}
\end{equation}
The integral over the imaginary part of the total Hamiltonian does not
necessarily produce zero even for square integrable wave functions, which is
rather typical of nonhomogeneous modifications. This should not come as a
surprise for $H_{I}$ formally describes absorption as, for instance, in the
so-called ``optical'' potentials. Even if the terms that $h_{I}$ consists of
are total derivatives, their convergence may not be strong enough to
guarantee that they vanish on the boundary in the infinity. Such terms
however are rather likely to be excluded either by the requirement that $%
|E_{QM}|<\infty $, or by the same requirement applied to the
field-theoretical energy functional for the modification which is given by
the following integral 
\begin{eqnarray}
\lefteqn{E_{FT}=\int \,d^{3}x\left\{ \frac{\hbar ^{2}}{2m}\left[ \left( \vec{%
\nabla}R\right) ^{2}+R^{2}\left( \vec{\nabla}S\right) ^{2}\right]
+c_{1}(\Delta S)^{2}+c_{2}\Delta S\frac{\Delta R}{R}+\right. }  \nonumber \\
&&\left. c_{3}\Delta S\left( \frac{\vec{\nabla}R}{R}\right) ^{2}+c_{4}\left( 
\frac{\Delta R}{R}\right) ^{2}+c_{5}\frac{\Delta R}{R}\left( \frac{\vec{%
\nabla}R}{R}\right) ^{2}+c_{6}\left( \frac{\vec{\nabla}R}{R}\right)
^{4}+VR^{2}\right\} .  \label{19}
\end{eqnarray}
Therefore, most likely, they do not play any physically relevant role.
Nevertheless, since one cannot say this with complete certainty without a
strict mathematical proof, we leave these terms in all the expressions they
happen to appear. It is easy to see that, as opposed to the Schr\"{o}dinger
equation, these two energy functionals differ. In fact, as pointed out in 
\cite{Pusz1}, the difference in question is a common feature of nonlinear
modifications of the Schr\"{o}dinger equation that are not homogeneous in
the wave function. It is not clear yet whether the homogeneity of a modified
Schr\"{o}dinger equation is of some physical significance, however what
appears to be evident is that the lack of such is not inconsequential as it
causes a greater departure from the properties of the Schr\"{o}dinger
equation than is the case for some homogeneous modifications. Although $%
E_{QM}=E_{FT}$ does not hold for all such modifications, the homogeneity is
a prerequisite for this to occur \cite{Pusz1}.

A similar construction cannot be performed for a three-dimensional
space-time. It is ruled out by the requirement that the Lagrangian is a
scalar. The same applies to any other odd-dimensional space-time. In two
dimensions the only Galilean invariant Lagrangian terms\footnote{%
Even if $\Delta S$ could, in principle, appear in the equations of motion,
it would have to come from a Lagrangian term $\left( \vec{\nabla}S\right)
^{2}$ which is not Galilean invariant. Postulating, as we do here, that the
Lagrangian be Galilean invariant is stronger than the mere Galilean
invariance of equations of motion.} that can contribute to the equations of
motion are $\Delta \ln aR^{2}$ and $(\vec{\nabla}lnaR^{2})^{2}.$ It is only
in the actual four-dimensional space-time that a cornucopia of new terms
emerges, some of them including the phase of the wave function in an
explicit manner. In higher dimensions the number of terms that preserve the
Galilean invariance of the Schr\"{o}dinger equation and do not introduce
dimensional constants becomes considerably greater.

The modification in question does not, in general, admit stationary states
for which $S=-Et/\hbar $ in an arbitrary potential that supports these
states in linear quantum mechanics. Such states, of energy $E$, are the only
stationary states allowed by the Schr\"{o}dinger equation. In the
hydrodynamic formulation they can be obtained from the stationarity
condition 
\begin{equation}
\frac{\partial R^{2}}{\partial t}=0.  \label{20}
\end{equation}
However, by demanding that such states exist for all potentials for which
they exist in linear quantum mechanics, one is able to limit the number of
free constants. To this end, let us notice that this condition applied to
(11) engenders that 
\begin{equation}
c_{2}\frac{\Delta R}{R}+c_{3}\left( \frac{\vec{\nabla}R}{R}\right) ^{2}=f,
\label{21}
\end{equation}
where $f$ is a harmonic function as it must satisfy $\Delta f=0$. This last
relation, notwithstanding a particular form of $f$, imposes a constraint on $%
R$ causing that (12) is satisfied only by a selected class of potentials $V$
and not all those that support stationary states in linear theory. To avoid
this limitation, one needs to put $c_{2}=c_{3}=0$. These coefficients have
to vanish also if the modified free Schr\"{o}dinger equation is to be
invariant under the time-reversal transformation as is the case for the
linear equation. Under these circumstances, the equation that determines the
energy of stationary states reads 
\begin{eqnarray}
\lefteqn{\frac{\hbar ^{2}}{m}\Delta R+2\left( E-V\right) R-2c_{4}\Delta
\left( \frac{\Delta R}{R^{2}}\right) -c_{5}\Delta \left[ \frac{1}{R}\left( 
\frac{\vec{\nabla}R}{R}\right) ^{2}\right] +\vec{\nabla}\cdot \left[
2c_{5}\left( \Delta R\frac{\vec{\nabla}R}{R^{3}}\right) \right. }  \nonumber
\\
&&\left. +4c_{6}\frac{1}{R}\left( \frac{\vec{\nabla}R}{R}\right) ^{3}\right]
+\frac{1}{R}\left[ 2c_{4}\left( \frac{\Delta R}{R}\right) ^{2}+3c_{5}\frac{%
\Delta R}{R}\left( \frac{\vec{\nabla}R}{R}\right) ^{2}+4c_{6}\left( \frac{%
\vec{\nabla}R}{R}\right) ^{4}\right] =0.  \label{22}
\end{eqnarray}
One further notes that the energy levels remain unmodified only if $%
c_{4}=c_{5}=c_{6}=0$, which clearly and uniquely singles out the original
Staruszkiewicz modification. However, the discussed stationary states may
not be the only stationary solutions of the modification. As a matter of
fact, as opposed to the Schr\"{o}dinger equation, the condition (20) leaves
room for more possibilities, that is, there may exists stationary solutions
more general than those for which the phase is uniquely determined as $%
S=-Et/\hbar $. With this condition put to work, they can be obtained by
solving eqs. (11-12).

In general, the only notable exception to this rule being the modification
of Bia\l ynicki-Birula and Mycielski, the nonlinear modifications of the
Schr\"{o}dinger equation do not have the standard classical limit in the
sense of the Ehrenfest theorem. The discussed modification is one of such
cases. The nonlinear terms it introduces lead to certain corrections to the
Ehrenfest relations which we will now work out. For a general observable $A$
one finds that 
\begin{equation}
\frac{d}{dt}\left\langle A\right\rangle =\frac{d}{dt}\left\langle
A\right\rangle _{L}+\frac{d}{dt}\left\langle A\right\rangle _{NL},
\label{23}
\end{equation}
where the nonlinear contribution due to $H_{NL}^{^{\prime }}=H_{R}^{^{\prime
}}+iH_{I}^{^{\prime }}$ can be expressed as 
\begin{equation}
\frac{d}{dt}\left\langle A\right\rangle _{NL}=\left\langle \left\{
A,H_{I}^{^{\prime }}\right\} \right\rangle -i\left\langle \left[
A,H_{R}^{^{\prime }}\right] \right\rangle .  \label{24}
\end{equation}
The brackets $<>$ denote the mean value of the quantity embraced, and $%
[\cdot ,\cdot ]$ and $\{\cdot ,\cdot \}$ denote commutators and
anticommutators, respectively. Specifying $A$ for the position and momentum
operators, one obtains the general form of the modified Ehrenfest relations 
\begin{equation}
m\frac{d}{dt}\left\langle \vec{r}\right\rangle =\left\langle \vec{p}%
\right\rangle +I_{1},  \label{25}
\end{equation}
\begin{equation}
\frac{d}{dt}\left\langle \vec{p}\right\rangle =-\left\langle \vec{\nabla}%
V\right\rangle +I_{2},  \label{26}
\end{equation}
where 
\begin{equation}
I_{1}=\frac{2m}{\hbar }\int \,d^{3}x\vec{r}R^{2}H_{I}^{^{\prime }},
\label{27}
\end{equation}
\begin{equation}
I_{2}=\int \,d^{3}xR^{2}\left( 2H_{I}^{^{\prime }}\vec{\nabla}S-\vec{\nabla}%
H_{R}^{^{\prime }}\right) -i\int d^{3}x\vec{\nabla}\left(
R^{2}H_{I}^{^{\prime }}\right) .  \label{28}
\end{equation}
In general, for nonhomogeneous modifications one cannot assume that $\int
d^{3}x\vec{\nabla}\left( R^{2}H_{I}^{^{\prime }}\right) =\frac{1}{2}\int
d^{3}x\vec{\nabla}h_{I}$ is zero exactly for the reason which prevented us
from discarding the imaginary terms $h_{I}$ in $E_{QM}$. However, the
integral in question is likely to vanish for all those configurations for
which $|E_{QM}|$ or $E_{FT}$ is finite. For the modification under study,
one can present integrals (27-28) in a more explicit form. Nevertheless, to
avoid rather cumbersome formulas, it seems more reasonable to treat them on
a case-by-case basis. In particular, the classical limit of the original
Staruszkiewicz modification will be considered in \cite{Pusz2}.

The Schr\"{o}dinger equation is invariant under the rescaling of the wave
function by an arbitrary complex constant $\lambda $, $\Psi ^{^{\prime
}}=\lambda \Psi $, which reflects the homogeneity of this equation. However,
as already noted, this is not the case for the modification in question. As
a result, the Staruszkiewicz modification does not allow for the weak
separability of composed systems \cite{Bial}. We will now discuss this issue
in greater detail.

The separability of the Schr\"{o}dinger equation is certainly an important
physical property and as such one would like to have it preserved in any
modification of this equation. Unfortunately, in general, this property
cannot be maintained in nonlinear wave mechanics, which means that in a
system consisting of two particles the very existence of one of them affects
the wave function of the other one. Thus, even in the absence of forces, the
rest of the world influences the behavior of an isolated particle. As
emphasized in \cite{Bial}, this is not necessarily a physically sound
situation. The Staruszkiewicz modification and its extended version
presented here violate this property. In this respect, they are similar to
another well known modification usually refered to as the cubic
Schr\"{o}dinger equation since it extends this equation by the term $|\Psi
|^{2}\Psi $. This is also what differs these modifications from some other
notable modifications of the Schr\"{o}dinger equation, to name only the most
studied ones, the modification of Bia\l ynicki-Birula and Mycielski \cite
{Bial} and the Doebner-Goldin modification \cite{Doeb}. To see how this
violation comes about, let us start by demonstrating the separability of
composed systems for the Schr\"{o}dinger equation in the hydrodynamic
formulation.

We are considering a quantum system made up of two noninteracting subsystems
in the sense that 
\begin{equation}
V(\vec{x}_{1},\vec{x}_{2},t)=V_{1}(\vec{x}_{1},t)+V_{2}(\vec{x}_{2},t).
\label{29}
\end{equation}
We will show that a solution of the Schr\"{o}dinger equation for this system
can be put in the form of the product of wave functions for individual
subsystems for any $t>0$, that is, $\Psi (x_{1},x_{2},t)=\Psi
_{1}(x_{1},t)\Psi _{2}(x_{2},t)=R_{1}(x_{1},t)R_{2}(x_{2},t)exp\left\{
i(S_{1}(x_{1},t)+S_{2}(x_{2},t))\right\} $ and that this form entails the
separability of the subsystems. The essential element here is that the
subsystems are initially uncorrelated which is expressed by the fact that
the total wave function is the product of $\Psi _{1}(\vec{x}_{1},t)$ and $%
\Psi _{2}(\vec{x}_{2},t)$ at $t=0$. What we will show then is that the
subsystems remain uncorrelated during the evolution and that, at the same
time, they also remain separated. It is the additive form of the total
potential that guarantees that no interaction between the subsystems occurs,
ensuring that they remain uncorrelated during the evolution. However, such
an interaction may, in principle, occur in nonlinear modifications of the
Schr\"{o}dinger equation even if the form of the potential itself does not
imply that. This is due to a coupling that a nonlinear term usually causes
between $\Psi _{1}(\vec{x}_{1},t)$ and $\Psi _{2}(\vec{x}_{2},t)$. The
Schr\"{o}dinger equation for the total system, assuming that the subsystems
have the same mass $m$, reads now

\begin{eqnarray}
\lefteqn{\hbar \frac{\partial R_{1}^{2}R_{2}^{2}}{\partial t}+\frac{\hbar
^{2}}{m}\left\{ \left( \vec{\nabla}_{1}+\vec{\nabla}_{2}\right) \cdot \left[
R_{1}^{2}R_{2}^{2}\left( \vec{\nabla}_{1}S_{1}+\vec{\nabla}_{2}S_{2}\right)
\right] \right\} =\hbar R_{2}^{2}\frac{\partial R_{1}^{2}}{\partial t}+\hbar
R_{1}^{2}\frac{\partial R_{2}^{2}}{\partial t}}  \nonumber \\
&&+\frac{\hbar ^{2}}{m}R_{2}^{2}\vec{\nabla}_{1}\cdot \left( R_{1}^{2}\vec{%
\nabla}_{1}S_{1}\right) +\frac{\hbar ^{2}}{m}R_{1}^{2}\vec{\nabla}_{2}\cdot
\left( R_{2}^{2}\vec{\nabla}_{2}S_{2}\right) =R_{1}^{2}R_{2}^{2}\left\{
\left[ \hbar \frac{1}{R_{1}^{2}}\frac{\partial R_{1}^{2}}{\partial t}%
+\right. \right.  \nonumber \\
&&\left. \left. \frac{\hbar ^{2}}{m}\frac{1}{R_{1}^{2}}\vec{\nabla}_{1}\cdot
\left( R_{1}^{2}\vec{\nabla}_{1}S_{1}\right) \right] +\left[ \hbar \frac{1}{%
R_{2}^{2}}\frac{\partial R_{2}^{2}}{\partial t}++\frac{\hbar ^{2}}{m}\frac{1%
}{R_{2}^{2}}\vec{\nabla}_{2}\cdot \left( R_{2}^{2}\vec{\nabla}%
_{2}S_{2}\right) \right] \right\} =0  \label{30}
\end{eqnarray}

and

\begin{eqnarray}
\lefteqn{\frac{\hbar ^{2}}{m}\left( \Delta _{1}+\Delta _{2}\right)
R_{1}R_{2}-2\hbar R_{1}R_{2}\frac{\partial (S_{1}+S_{2})}{\partial t}-\frac{%
\hbar ^{2}}{m}R_{1}R_{2}\left( \vec{\nabla}_{1}S_{1}+\vec{\nabla}%
_{2}S_{2}\right) ^{2}-}  \nonumber \\
&&\left( V_{1}+V_{2}\right) R_{1}R_{2}=\frac{\hbar ^{2}}{m}R_{2}\Delta
_{1}R_{1}+\frac{\hbar ^{2}}{m}R_{1}\Delta _{2}R_{2}-2\hbar R_{1}R_{2}\frac{%
\partial S_{1}}{\partial t}-2\hbar R_{1}R_{2}\frac{\partial S_{2}}{\partial t%
}  \nonumber \\
&&+\frac{\hbar ^{2}}{m}R_{1}R_{2}\left( \vec{\nabla}_{1}S_{1}\right) ^{2}+%
\frac{\hbar ^{2}}{m}R_{1}R_{2}\left( \vec{\nabla}_{2}S_{2}\right)
^{2}-V_{1}R_{1}R_{2}-V_{2}R_{1}R_{2}=  \nonumber \\
&&R_{1}R_{2}\left\{ \left[ \frac{\hbar ^{2}}{m}\frac{\Delta _{1}R_{1}}{R_{1}}%
-2\hbar \frac{\partial S_{1}}{\partial t}+\frac{\hbar ^{2}}{m}\left( \vec{%
\nabla}_{1}S_{1}\right) ^{2}-V_{1}\right] +\left[ \frac{\hbar ^{2}}{m}\frac{%
\Delta _{2}R_{2}}{R_{2}}-2\hbar \frac{\partial S_{2}}{\partial t}+\right.
\right.  \nonumber \\
&&\left. \left. \frac{\hbar ^{2}}{m}\left( \vec{\nabla}_{2}S_{2}\right)
^{2}-V_{2}\right] \right\} =0.  \label{31}
\end{eqnarray}
Implicit in the derivation of these equations is the fact that $\vec{\nabla}%
_{1}f_{1}\cdot \vec{\nabla}_{2}g_{2}=0$, where $f_{1}$ and $g_{2}$ are
certain scalar functions defined on the configuration space of particle 1
and 2, correspondingly. What we have obtained is a system of two equations,
each consisting of terms (in square brackets) that pertain to only one of
the subsystems. By dividing the first equation by $R_{1}^{2}R_{2}^{2}$ and
the second one by $R_{1}R_{2}$, one completes the separation of the
Schr\"{o}dinger equation for the compound system into the equations for the
subsystems. Moreover, we have also showed that indeed the product of wave
functions of the subsystems evolves as the wave function of the total system.

This is, however, not so for the Staruszkiewicz modification which is
particularly easy to demonstrate for its original formulation. In this case
the continuity equation for our composite system reads 
\begin{equation}
\hbar \frac{\partial R_{1}^{2}R_{2}^{2}}{\partial t}+\frac{\hbar ^{2}}{m}%
\left\{ \left( \vec{\nabla}_{1}+\vec{\nabla}_{2}\right) \cdot \left[
R_{1}^{2}R_{2}^{2}\left( \vec{\nabla}_{1}S_{1}+\vec{\nabla}_{2}S_{2}\right)
\right] \right\} -c_{1}\left( \Delta _{1}^{2}+\Delta _{2}^{2}\right) \left(
S_{1}+S_{2}\right) =0.  \label{32}
\end{equation}
This can be brought to 
\begin{eqnarray}
R_{1}^{2}R_{2}^{2}\left\{ \left[ \hbar \frac{1}{R_{1}^{2}}\frac{\partial
R_{1}^{2}}{\partial t}+\frac{\hbar ^{2}}{m}\frac{1}{R_{1}^{2}}\vec{\nabla}%
_{1}\cdot \left( R_{1}^{2}\vec{\nabla}_{1}S_{1}\right) \right] +\right.  
\nonumber \\
\left. \left[ \hbar \frac{1}{R_{2}^{2}}\frac{\partial R_{2}^{2}}{\partial t}+%
\frac{\hbar ^{2}}{m}\frac{1}{R_{2}^{2}}\vec{\nabla}_{2}\cdot \left( R_{2}^{2}%
\vec{\nabla}_{2}S_{2}\right) \right] -c_{1}\left[ \frac{\Delta _{1}^{2}S_{1}%
}{R_{1}^{2}R_{2}^{2}}+\frac{\Delta _{2}^{2}S_{2}}{R_{1}^{2}R_{2}^{2}}\right]
\right\}  &=&0  \label{33}
\end{eqnarray}
which clearly shows that the separation of subsystems cannot be achieved in
this situation due to the modification term. In the similar manner, one can
prove that the same holds for other terms of the discussed modification. 

The discussed separability is called the weak separability since it assumes
that the wave function of the total system is the product of the wave
functions of its subsystems in which it differs from the strong version of
separability that does not employ this assumption. As shown by L\"{u}cke 
\cite{Luc1, Luc2}, weakly separable modifications, such as the modification
of Bia\l ynicki-Birula \cite{Bial} or the Doebner-Goldin modification \cite
{Doeb}, can still violate separability when the compound wave function is
not factorizable, and thus they are not strongly separable. An alternative
effective approach to the strong separability has been proposed by Czachor 
\cite{Czach1}. This approach treats the density matrix as the basic object
subjected to the quantum equations of motion which are modified\footnote{%
What this means in practice is that the basic equation is the nonlinear von
Neumann equation \cite{Czach2} for the density matrix instead of some
nonlinear Schr\"{o}dinger equation for the pure state.} compared to a
nonlinear Schr\"{o}dinger equation for the pure state. It admits a large
class of nonlinear modifications including those ruled out by the
fundamentalist approach advocated by L\"{u}cke and even those that are not
weakly separable as, for instance, the cubic nonlinear Schr\"{o}dinger
equation. As asserted by Czachor \cite{Czach1}, the modification of this
type should be separable in the strong effective approach.

However, the presented analysis of separability of the extended
Staruszkiewicz modification shows that this modification is not separable in
the fundamentalist approach. This is not unlike the Bia\l ynicki-Birula and
Mycielski modification but for a different reason. What excludes the
Staruszkiewicz model of nonlinearity is the fact that its equations do not
permit a weakly separable multi-particle extension that would comply with
the strong separability. This fact is ultimately attributed to the
nonhomogeneity of the model. Let us demonstrate this on the example of the
cubic nonlinear Schr\"{o}dinger equation that shares this property with the
modification under study. As already noted, the cubic nonlinearity is not
weakly separable, yet one can  find a weakly separable multi-particle
extension for it. For instance, for two particles the following extension is
weakly separable\footnote{%
Of course, this extension represents a different theory from the
straightforward extension that is not weakly separable for it introduces the
coupling $\left| \Psi _{1}\right| ^{2}\left| \Psi _{2}\right| ^{2}$, but, in 
principle, there is no reason to dismiss it. } 
\begin{equation}
i\hbar \frac{\partial \left( \Psi _{1}\Psi _{2}\right) }{\partial t}=\left[
-\left( \frac{\hbar ^{2}\Delta _{1}}{2m_{1}}+\frac{\hbar ^{2}\Delta _{2}}{%
2m_{2}}\right) +\left| \Psi _{1}\right| ^{2}+\left| \Psi _{2}\right|
^{2}\right] \Psi _{1}\Psi _{2}.  \label{34}
\end{equation}
Unfortunately, as this example demonstrates, the only way to make a
nonhomogeneous modification weakly separable is by splitting the total wave
function of the system into the wave functions of its constituents.
Obviously, this cannot be extended to nonfactorizable compound wave
functions. Such an extension is, however, possible for the modification
presented in \cite{Pusz3} which, being weakly nonseparable, is homogeneous.
This is possible because this particular weakly separable extension does not
result in splitting the compound wave function. 

Even if it may appear to be relatively easy to propose a nonlinear
modification of the Schr\"{o}dinger's equation, it is more difficult to come
up with a wide array of physically sensible assumptions the modifications of
this equation should satisfy. Bia\l ynicki-Birula and Mycielski \cite{Bial}
were first to emphasize the physical significance of the postulate of
separability of composite systems. Other attempts at modifying this
equation, mentioned above, satisfy this postulate as well. Nevertheless, it
is not entirely clear as to what extent one can consider components of a
composite quantum-mechanical system to be separable and thus what status one
should ascribe to the postulate in question. Being a natural feature of the
Schr\"{o}dinger equation, it may not necessarily reflect any physical
property of quantum-mechanical reality, in the same sense as the equation
itself does not account for the collapse of the wave function.

Since it seems nontrivial to construct a modification of Schr\"{o}dinger's
equation implementing the discussed feature there may be only a limited
class of modifications of this kind. In this respect, the postulate of
separability offers a certain way to restrict the number of all conceivable
modifications. One can also use it to classify the modifications into those
which meet it and the remaining ones. The modifications from the latter
category would be deemed physically attractive by arguments of a different
kind, be it of simplicity or integrability, as is the case for the cubic
Schr\"{o}dinger equation, or of some particular or desirable properties. On
the other hand, if one modifies the Schr\"{o}dinger equation in an attempt
to incorporate the collapse of wave function into it, the discussed
postulate may be overly restrictive. It is plausible that without a coupling
with the rest of the world the collapse of wave function of an individual
system may never take place. Moreover, it is not out of the question that to
solve the problem of collapse by introducing such a nonlinear coupling one
needs to give up all features of the linear Schr\"{o}dinger equation and not
only some of them. Therefore, it seems to be justifiable to study also these
modifications of this equation that do not satisfy the assumption of
separability, especially if they possess certain attractive, unique features
as the modification discussed here.

It is not clear how to interpret the new terms of Lagrangian (8) or (9) and
what their physical significance is. In fact, this plethora of terms can be
slightly discouraging as it is now in sharp contrast with the original
Staruszkiewicz modification involving only one dimensionless constant. It
was this uniqueness of Staruszkiewicz's modification that would give it some
advantage over other extensions of the fundamental equation of quantum
mechanics. One should however be stressed that this situation is quite
typical for modifications of the Schr\"{o}dinger equation. The
Doebner-Goldin modification introduces exactly the same number of terms,
although not all of them are allowed if its equations are to be Galilean
invariant, and the number of terms in the Weinberg modification does not
seem to be limited by any physical principle as well. The simplest of them,
the modification of Bia\l ynicki-Birula and Mycielski leads to only one new
physically relevant constant, which is solely due to a certain simplifying
condition. This condition carries no physical contents, though. We propose
to call the dimensionless parameters of this extended modification the
Staruszkiewicz quantum numbers by analogy to hydrodynamical Reynolds
numbers, in part because the hydrodynamic formulation seems to be the most
convenient to study the modification discussed and in part because of their
dimensionless nature. The simplest solution to this problem of abundance of
terms is to subject them to a test of physical plausibility which, as
demonstrated above, only the term suggested by Staruszkiewicz can pass.
Other terms would necessarily change energy levels of quantum-mechanical
systems. While one can probably make these changes arbitrarily small by
imposing appropriate constraints on the Staruszkiewicz quantum numbers, it
is certainly more elegant to choose another option which consist in admiting
only the terms that do not affect the stationary states of
quantum-mechanical systems at all. It is only the original Staruszkiewicz
model that meets this requirement. In this respect, the modification
proposed by Staruszkiewicz differs from the proposals of Bia\l ynicki-Birula
and Mycielski, Weinberg, and Doebner-Goldin, as all of them imply changes in
energy levels of known quantum systems by some finite, measurable, at least
in principle, values.

In any case, it is the experiment that should decide whether the original
Staruszkiewicz modification or its extension proposed here reflect any
aspect of reality. In this context, it would be interesting to see if the
methods used to set bounds on the nonlinear parameters in the Weinberg or
Bia\l ynicki-Birula and Mycielski modification could be of any use here. The
same question applies to the method of neutron interference proposed by
Shimony \cite{Shim} and subsequently implemented to test the latter
modification \cite{Shull}. This important point of our proposal is left for
further studies.

\section*{Acknowledgments}

I would like to thank Professor Pawe{\l } O. Mazur for bringing my attention
to the work of Professor Staruszkiewicz that started my interest in
nonlinear modifications of the Schr\"{o}dinger equation. I am very indebted
to Professor Andrzej Staruszkiewicz for a critical reading of the
preliminary version of this paper and discussions in the summer of 1996 and
1998. A correspondence with Professor Wolfgang L\"{u}cke concerning the 
problem of separability in nonlinear quantum mechanics and a correspondence 
with Dr. Marek Czachor about many issues of nonlinear quantum mechanics are 
also gratefully acknowledged. This work was partially supported by the NSF 
grant No. 13020 F167 and the ONR grant R\&T No. 3124141.

\end{document}